\documentclass[aps,pra,twocolumn,amsmath,amssymb,preprintnumbers,showpacs]{revtex4}
\usepackage{graphicx}
\usepackage{dcolumn}

\begin{document}


\title{Measurement-induced manipulation of the quantum-classical border}

\author{Sabrina Maniscalco}
\affiliation{Department of Physics, University of Turku, FIN-20014
Turun yliopisto, Turku, Finland}
\author{Jyrki Piilo}
\affiliation{Department of Physics, University of Turku, FIN-20014
Turun yliopisto, Turku, Finland}
\author{Kalle-Antti Suominen}
\affiliation{Department of Physics, University of Turku, FIN-20014
Turun yliopisto, Turku, Finland}

\email{sabrina.maniscalco@utu.fi}

\date{\today}

\begin{abstract}
We demonstrate the possibility of controlling the border between
the quantum and the classical world by performing nonselective
measurements on quantum systems. We consider a quantum harmonic
oscillator initially prepared in a Schr\"{o}dinger cat state and
interacting with its environment. We show that the environment
induced decoherence transforming the cat state into a statistical
mixture can be strongly inhibited by means of appropriate
sequences of measurements.
\end{abstract}

\maketitle

\section{Introduction}
The increasing ability in the coherent control and manipulation of
the state of quantum systems has paved the way to experiments able
to monitor the transition from quantum superpositions, such as the
paradigmatic Schr\"{o}dinger cat states, to classical statistical
mixtures\cite{engineerNIST}. The emergence of the classical world
from the quantum world, due to decoherence induced by the
environment, has been extensively investigated in the last few
decades both in connection to fundamental issues of quantum theory
and in relation to the emerging quantum technologies. The fragile
nature of quantum superpositions and entangled states exploited,
e.g., in quantum communication, quantum computation, and quantum
metrology, makes these potentially very powerful techniques also
very delicate. For this reason several methods have been proposed
in order to protect quantum states from decoherence and
dissipation. For example, methods based on decoherence free
subspaces, dynamical decoupling and bang-bang techniques, just to
mention a few, have been investigated~\cite{DFS}. Recently the
connection between these techniques and the quantum Zeno effect
has been clarified~\cite{Facchi05}.

In a recent letter we have studied the conditions for observing
the Zeno and anti-Zeno effects in a damped harmonic
oscillator~\cite{Maniscalco06a}. The quantum Zeno and anti-Zeno
effects~\cite{Misra,Lane} predict the inhibition and the
enhancement, respectively, of the decay of the initial state due
to a series of measurements aimed at checking whether the system
is still in its initial state or not~\cite{FacchiRev}. Typically,
when studying Zeno and anti-Zeno dynamics, the system is assumed
to be initially prepared in an eigenstate of the free Hamiltonian,
e.g., in our case, a Fock state. This is the situation we have
considered in Refs.~\cite{Maniscalco06a,Piilo07a}. The aim of this
paper is to see whether the quantum Zeno effect can be exploited
also to inhibit quantum decoherence when the system is initially
prepared in a Schr\"{o}dinger cat state. The analysis of Zeno and
anti-Zeno phenomena in the context of the damped harmonic
oscillator gives us the possibility of exploring the modification
of the quantum-classical transition as an effect of measurements
performed on the system. This possibility stems from the fact that
the harmonic oscillator possesses both quantum states, such as
Fock states and superposition of coherent states, and classical
(or semiclassical) states, such as the coherent and the thermal
states.

Another aspect discussed in the paper is the connection between
the dynamics in presence of non-selective energy measurements and
the dynamics in presence of modulation of the system-reservoir
coupling constant. The second scenario may be useful in
implementing experiments aimed at revealing the quantum Zeno and
anti-Zeno effects with engineered reservoirs~\cite{engineerNIST}.
We will show that a simple periodic modulation of the
system-reservoir coupling constant is equivalent to performing
non-selective energy measurements, as one would expect from the
results presented in Ref.~\cite{Facchi05}. An experimental
verification of the Zeno and anti-Zeno effects with engineered
reservoirs would allow one to observe these phenomena through
indirect measurements, contrarily to the direct ones of
Ref.~\cite{RaizenZeno}, in the spirit of the \lq
genuine\rq~quantum Zeno effect~\cite{Koshino05}.

\section{Quantum Zeno and anti-Zeno effects for the damped harmonic oscillator}
We consider a harmonic oscillator linearly coupled with a
reservoir modelled as an infinite chain of non-interacting
oscillators~\cite{Weiss99a,Breuer02a,Feynman63a,Caldeira83a,Haake85a,Grabert88a,Hu92a}.
The dynamics of the damped harmonic oscillator is described, in
the secular approximation and in the interaction picture, by means
of the following generalized master equation
\cite{Intravaia03a,Maniscalco04b}
\begin{eqnarray}
&&\frac{ d \rho(t)}{d t}= \frac{\Delta(t) \!+\! \gamma (t)}{2}
\left[2 a \rho(t) a^{\dag}- a^{\dag} a \rho(t)  - \rho(t) a^{\dag}
a \right]
\nonumber \\
&& +\frac{\Delta(t) \!-\! \gamma (t)}{2} \left[2 a^{\dag} \rho(t)
a - a a^{\dag} \rho(t) - \rho(t) a a^{\dag}
 \right]. \nonumber \\
 \label{Eq:MQbm}
\end{eqnarray}
In this equation, $a$ and $a^{\dag}$ are the annihilation and
creation operators, and $ \rho(t)$ is the reduced density matrix
of the system oscillator. It is worth noticing that the only
approximation done in the derivation of the master equation
(\ref{Eq:MQbm}) is the secular approximation typical of quantum
optical systems. However, no Born-Markov approximation has been
done, therefore this master equation describes the non-Markovian
dynamics of the system. For times much longer than the reservoir
correlation time $\tau_R$, the time dependent coefficients
$\Delta(t)+\gamma(t)$ and $\Delta(t)- \gamma(t)$ approach their
stationary values
\begin{eqnarray}
\gamma_{1}^M &\equiv& \Gamma [N(\omega_0)+1], \label{eq:marg1}\\
\gamma_{-1}^M &\equiv& \Gamma N(\omega_0), \label{eq:margm1}
\end{eqnarray}
respectively, [See Appendix A of Ref.~\cite{Maniscalco04b} for
details] and the master equation (\ref{Eq:MQbm}) reduces to the
well known Markovian master equation for the damped harmonic
oscillator.

In contrast to other non-Markovian dynamical systems, the master
equation~(\ref{Eq:MQbm}) is local in time, i.e.~it does not
contain memory integrals. All the non-Markovian character of the
system is contained in the time dependent coefficients appearing
in the master equation. These coefficients, namely the diffusion
coefficient $\Delta(t)$ and the damping coefficient $\gamma(t)$,
depend uniquely on the form of the reservoir spectral density. To
second order in the dimensionless system-reservoir coupling
constant $g$, they take the form \cite{Hu92a,Maniscalco04b}
\begin{eqnarray}
\!\Delta(t)\!\!\!&=&\! \!g^2\!\! \!\!\int_0^t \!\!\!\!
\int_0^{\infty}\!\!\!\!\! \!\!d \omega d t_1 J(\!\omega\!)\!
\left[2N(\omega)+1\right] \!\cos(\omega t_1\!) \!\cos (\omega_0
t_1\!),
\label{delta} \\
\gamma(t)\!\!&=& g^2 \int_0^t\!\! \int_0^{\infty}\!\!\! d \omega d
t_1  J(\omega) \sin(\omega t_1) \sin (\omega_0 t_1), \label{gamma}
\end{eqnarray}
with $J(\omega)$ the reservoir spectral density, $N(\omega) =
(e^{\hbar \omega/k_B T}-1)^{-1}$ the average number of reservoir
thermal photons, $k_B$ the Boltzmann  constant, and $T$ the
reservoir temperature. We note that, for high $T$, i.e. $N(\omega)
\gg 1$, $\Delta(t)\gg \gamma(t)$.

\subsection{Evolution in presence of non-selective energy measurements}

The time evolution of this system has been extensively
studied~\cite{Feynman63a,Caldeira83a,Haake85a,Grabert88a,Hu92a,Intravaia03a,Maniscalco04b,Zurek,Anglin96,Maniscalco04a,EPJRWA},
and in particular the dynamics of initial nonclassical states due
to environment induced decoherence has been discussed in various
regimes. Here we describe the modifications of the open system
dynamics due to a series of non-selective energy measurements,
described in terms of the projection operator $\hat{P}$
\begin{eqnarray}
\hat{P} \rho = \sum_n P_n \vert n \rangle \langle n \vert,
\label{eq:Pmeas}
\end{eqnarray}
where $\vert n \rangle$ are the Fock states of the harmonic
oscillator and $P_n = \langle n \vert \rho \vert n \rangle $ are
the diagonal elements of the reduced density matrix. Essentially
the effect of these measurements is to erase instantaneously all
the coherences, without selecting any of the energy states of the
systems. We assume that during the time evolution the harmonic
oscillator is subjected to a series of non-selective energy
measurements, and we indicate with $\tau$ the time interval
between two successive measurements. We assume that $\tau$ is so
short (and/or the coupling so weak) that second order processes
may be neglected.

Following the derivation given, for a generic system, in
Ref.~\cite{Facchi05} we can write down a coarse grained master
equation governing the system time evolution in presence of $m$
nonselective measurements
\begin{eqnarray}
\frac{ d \rho(t)}{d t} &=& \gamma_1(\tau) \hat{P} \left[ a \hat{P}
\rho (t) a^{\dagger} - \frac{1}{2} a^{\dagger}a \hat{P} \rho(t) -
\frac{1}{2} \hat {P} \rho(t) a^{\dagger}a\right]
\nonumber \\
&+& \gamma_{-1}(\tau) \hat{P} \left[  a^{\dagger} \hat{P} \rho (t)
a - \frac{1}{2}a a^{\dagger}\hat{P}\rho(t)
- \frac{1}{2} \hat{P}\rho(t) a a^{\dagger}\right]. \nonumber \\
&& \label{Eq:RecME}
\end{eqnarray}
Here $t=m\tau$ and $\gamma_{\pm 1}(\tau) $ are given by
\begin{eqnarray}
\gamma_{\pm 1}(\tau)= \tau \int_{-\infty}^{\infty}d \omega
\kappa^{\beta}(\omega) {\rm sinc}^2 \left( \frac{\omega\mp
\omega_0}{2}\tau \right), \label{eq:gammapmtau}
\end{eqnarray}
where ${\rm sinc} (x) = \sin{x}/x $, and the thermal spectral
density $\kappa^{\beta}(\omega)$ is defined as
\begin{eqnarray}
\kappa^{\beta} (\omega)= J (\omega) \theta(\omega)
[N(\omega)+1]+J(- \omega) \theta (-\omega) N(- \omega),
\label{eq:thermalk}
\end{eqnarray}
with $\theta(\omega)$ the unit step function.

The dynamics described by the master equation~(\ref{Eq:RecME}) is
such that only the diagonal elements of the density matrix are
nonzero, due to the effect of the nonselective measurement
described by Eq.~(\ref{eq:Pmeas}). The time evolution of the
number probability distribution $P_n(t)= \langle n \vert \rho(t)
\vert n \rangle$ is easily obtained by Eq.~(\ref{Eq:RecME}), and
reads as
\begin{eqnarray}
\dot{P}_n(t)&=& \gamma_1(\tau) \left[(n+1)P_{n+1}(t)- n
P_n(t)\right] \nonumber
\\
& +& \gamma_{-1} (\tau) \left[n P_{n-1}(t)-(n+1)P_n (t)\right].
\label{eq:Pn}
\end{eqnarray}
We note that the decay rates $\gamma_1(\tau)$ and
$\gamma_{-1}(\tau)$ do not depend on time $t$, hence these rate
equations are formally equivalent to those obtained from the
Markovian master equation for the damped harmonic oscillator,
provided that one identifies $ \gamma_1(\tau)$ with $\Gamma
[N(\omega_0)+1]$ and $\gamma_{-1}(\tau)$ with $\Gamma N(\omega)$.
The effect of the nonselective energy measurements is therefore
twofold. On the one hand they destroy the off diagonal elements of
the density matrix, and on the other hand they modify the decay
coefficients appearing in the rate equations in a way which
depends crucially both on the system/reservoir parameters and on
the interval $\tau$ between the measurements. In order to
understand in more detail how the decay coefficients are modified
when compared to the Markovian ones we further investigate
$\gamma_{\pm 1}(\tau)$, as given by Eq.~(\ref{eq:gammapmtau}) with
Eq.~(\ref{eq:thermalk}). Recasting
Eqs.~(\ref{delta})-(\ref{gamma}) as
\begin{eqnarray}
\!\!\Delta(t)&=& \!\!\frac{t}{2}\int_0^{\infty} \!\!\!\!d \omega
J(\omega) \left[ N(\omega)+\frac{1}{2}\right] \Big\{ {\rm sinc}
[(\omega-\omega_0)t] \nonumber \\
&+& {\rm sinc} [(\omega + \omega_0)t] \Big\}, \\
\!\!\!\!\!\!\gamma(t)\!\!&=&\!\! \frac{t}{2}\int_0^{\infty}
\!\!\!\!\!d \omega \frac{J(\omega)}{2}  \left\{ {\rm sinc}
[(\omega-\omega_0)t] - {\rm sinc} [(\omega + \omega_0)t] \right\}
\nonumber \\
\end{eqnarray}
and integrating the sum and the difference of these coefficients
over the time interval $\tau$, it is straightforward to prove that
\begin{eqnarray}
\gamma_{\pm 1}(\tau)=\frac{1}{\tau} \int_0^{\tau} dt
\left[\Delta(t)\pm\gamma(t)\right].
\end{eqnarray}
The equation above shows the connection between the coefficients
of the non-Markovian master equation (\ref{Eq:MQbm}) and the
coefficients $\gamma_{\pm}(\tau)$ modified by the presence of the
nonselective energy measurements. We note first of all that when
the interval between the measurements $\tau$ is much greater than
the reservoir correlation time $ \tau_R$, $\gamma_{\pm 1}(\tau)
\simeq \gamma_{\pm 1}^M $, since $\Delta(t)$ and $\gamma(t)$
quickly set to their Markovian stationary values. In this case one
recovers for $P_n(t)$ the usual Markovian dynamics, i.e., the
presence of the nonselective energy measurements cannot modify the
Markovian decay of the number probability distribution. Stated
another way, in order to affect the dynamics one needs to perform
the measurements at time intervals shorter or of the same order of
the reservoir correlation time. This is well known in the theory
of the quantum Zeno effect, since such effect is crucially related
to the short time initial quadratic behavior of the survival
probability, i.e. of the probability that a system prepared in a
given initial state is still in that initial state after a time
$t$~\cite{FacchiRev}.

As we mentioned at the beginning of this section for high $T$
reservoirs $\Delta(t) \gg \gamma(t)$, therefore, for times much
smaller than the thermalization time $\tau_{\rm th} \simeq
1/\Gamma$, with $\Gamma$ defined in
Eqs.~(\ref{eq:marg1})-(\ref{eq:margm1}),
\begin{eqnarray}
\gamma_{1}(\tau) \simeq \gamma_{-1}(\tau)\simeq \frac{1}{\tau}
\int_0^{\tau} dt \Delta(t).
\end{eqnarray}
As we have noted and discussed in Ref.~\cite{Maniscalco06a}, for
high $T$ reservoirs, therefore, the modified decay rates depend
only on the diffusion coefficient $\Delta(t)$ describing
environment induced decoherence. In other words the repetition of
nonselective energy measurements at short intervals $\tau$ forces
the system to experience an enhanced or reduced environment
induced decoherence, depending on the form of the reservoir
spectrum, and hence on the temporal behavior of $\Delta(t)$.

Before discussing the effect of measurements on the decoherence of
an initial Schr\"{o}dinger cat state, we want to address the
connection between the dynamics described in this section and the
situation in which the measurements are replaced by a periodic
modulation in the system-reservoir coupling constant.

\section{Mimicking the effect of measurements with engineered reservoirs}
We consider here the case in which the coupling between the system
oscillator and the reservoir is weak enough to justify the use of
second order perturbation theory. Instead of introducing a series
of nonselective energy measurements we consider the case in which
the microscopic system reservoir coupling constant $g$ is
modulated as follows
\begin{eqnarray}
g (t) = \left\{
\begin{array}{ll}
g & {\rm for} \hspace{0.3 cm} m (\tau +  \delta \tau) \le t < m (\tau +  \delta \tau) + \tau, \\
0 & {\rm for} \hspace{0.3 cm} m (\tau +  \delta \tau )+ \tau \le t
< (m+1)(\tau+\delta \tau),
\end{array}
\right.
\end{eqnarray}
with $m=0,1,2,...$. Stated another way, the coupling between the
system and the reservoir is interrupted, for a short time $\delta
\tau \ll \tau$, periodically at intervals $\tau$. As we will show
in the following, under certain conditions, these short
interruptions mimic the nonselective energy measurements. We refer
to the system described here with the name of shuttered reservoir,
according to the terminology we have used in Ref.~\cite{Piilo07a}.
We will assume in the following that the time intervals $\delta
\tau$ are so small that the free evolution of the system can be
neglected. This assumption is not necessary when we deal with the
dynamics of an initial Fock state, as we have explained in
Ref.~\cite{Piilo07a}.

\subsection{The recursive solution}
Since both $\Delta (t)$ and $\gamma(t)$ are proportional to the
coupling constant $g$ [See Eqs.~(\ref{delta})-(\ref{gamma})], we
can solve the dynamics using recursively the solution of the
master equation Eq.~(\ref{Eq:MQbm}). More in detail, we solve the
master equation given by Eq.~(\ref{Eq:MQbm}), e.g. in terms of the
quantum characteristic function (QCF) as done in
Ref.~\cite{Intravaia03a}, and we use the solution at time $\tau$
as initial condition at time $\tau +\delta \tau$. We then
calculate the solution at time $2 \tau + \delta \tau$ and use this
as initial condition once more. In this way we obtain the
following coarse grained recursive solution for the QCF,
\begin{eqnarray}
\chi_{m,\tau} (\xi) = \exp \left[-f_m(\tau)|\xi|^2 \right] \chi_0
\left(e^{-m\Gamma(\tau) \xi /2} \right), \label{eq:recsol}
\end{eqnarray}
where $m$ is the number of interruptions in the system-reservoir
coupling, $\chi_0(\xi)$ is the QCF of the initial state, and
\begin{eqnarray}
f_m(\tau)&=& \Delta_{\Gamma}(\tau) \frac{1-e^{-m
\Gamma(\tau)}}{1-e^{-\Gamma(\tau)}},
\end{eqnarray}
with
\begin{eqnarray}
\Gamma(\tau)&=& 2\int_0^{\tau} dt \: \gamma(t), \label{Gamma}
\label{Eq:Gamma}\\
\Delta_{\Gamma}(\tau) &=& e^{-\Gamma(\tau)}\int_0^{\tau}dt \:
e^{\Gamma(t)}\Delta(t) \label{DeltaGamma}. \label{Eq:DG}
\end{eqnarray}

\subsection{Comparison between the shuttered reservoir and the nonselective measurements scenarios}
It is straightforward to demonstrate that the density matrix
corresponding to the QCF solution given by Eq.~(\ref{eq:recsol})
is the solution of the following master equation
\begin{eqnarray}
\frac{ d \rho(t)}{d t} &=& \gamma_1(\tau) \left[ a  \rho (t)
a^{\dagger} - \frac{1}{2} a^{\dagger}a  \rho(t) - \frac{1}{2}
\rho(t) a^{\dagger}a\right]
\nonumber \\
&+& \gamma_{-1}(\tau)  \left[  a^{\dagger}  \rho (t) a -
\frac{1}{2}a a^{\dagger}\rho(t)
- \frac{1}{2} \rho(t) a a^{\dagger}\right]. \nonumber \\
&& \label{Eq:RecME2}
\end{eqnarray}
In order to prove it one only needs to transform the master
equation above into a partial differential equation for the QCF
[See, e.g., Appendix 12 of Ref.~\cite{Barnett}], and then to
verify by direct substitution that Eq.~(\ref{eq:recsol}) satisfies
the partial differential equation.

Comparing Eq.~(\ref{Eq:RecME2}), describing the dynamics in
presence of a shuttered reservoir, with Eq.~(\ref{Eq:RecME}),
describing the dynamics in presence of nonselective energy
measurements, one notices immediately that the difference between
the two physical situations consists in the fact that while the
nonselective measurements always set to zero the coherences, in
the case of a shuttered reservoir the off diagonal elements of the
density matrix do not vanish. The rate equations for the number
probability distribution $P_n(t)$, however, do coincide, and are
given in both cases by Eq.~(\ref{eq:Pn}). This leads us to two
conclusions.

Firstly, whenever the initial state of the system is a Fock state,
or any state which is diagonal in the Fock state basis, the system
dynamics in presence of shuttered reservoirs coincides exactly
with the dynamics in presence of nonselective energy measurements,
since for this type of initial condition the density matrix, for
the shuttered reservoir case, remains diagonal at all times $t$.
Therefore, in this case, the shuttered reservoir mimics the
Zeno/anti-Zeno dynamics in presence of nonselective energy
measurements. This is interesting because it may be easier to
realize experimentally a shuttered reservoir, instead of a
sequence of nonselective energy measurements, by using reservoir
engineering techniques as those used in the trapped ions
context~\cite{engineerNIST}. In
Refs.~\cite{Maniscalco06a,Piilo07a} we briefly discuss a possible
implementation of the shuttered reservoir with trapped ions.

Secondly, the time evolution of the number probability
distribution, and therefore of all the observables which are
diagonal in the Fock state basis, coincides for the two scenarios
discussed in this paper. In Sec.~\ref{sec:Pn} we will examine in
detail the dynamics of the number probability distribution for the
case of an initial Schr\"{o}dinger cat state, and we will show
that for certain values of the parameters one can manipulate the
quantum-classical border prolonging or shortening the \lq
life\rq~of the cat.

\section{Control of the quantum-classical border}
In this section we consider the case in which the harmonic
oscillator is initially prepared in a Schr\"{o}dinger cat state of
the form
\begin{eqnarray}
\vert \Psi \rangle = \frac{1}{\sqrt{{\cal N}}} \left( \vert \alpha
\rangle + \vert - \alpha \rangle \right), \label{eq:inistate}
\end{eqnarray}
where $\vert \alpha \rangle$, with $\alpha \in \mathbb{R}$ is a
coherent state and
\begin{eqnarray}
{\cal N}^{-1}=2 \left[1+ \exp \left(-2 |\alpha|^2 \right) \right].
\end{eqnarray}
This state is also known as even coherent state due to the fact
that only the even components of the number probability
distribution are nonzero. These oscillations in the number state
probability are a strong sign of the nonclassicality of this
state. This and other nonclassical properties of the even coherent
state, such as the negativity of the corresponding Wigner
function, have been extensively studied in the literature [See,
e.g., Ref.~\cite{Buzek92} and references therein]. In particular,
the decoherence and dissipation due to the interaction with the
environment, in the Markovian approximation, have been studied in
Ref.~\cite{Buzek92}, and the non-Markovian dynamics has been
discussed in Refs.~\cite{Zurek,Maniscalco04b}. In Fig.~1 we show
the number probability distribution and the Wigner function for
the even coherent state. This state has been realized in the
trapped ion context and the transition from a quantum
superposition to a classical statistical mixture has been observed
experimentally~\cite{engineerNIST}.

\begin{figure}
\centering
\includegraphics[scale=0.4]{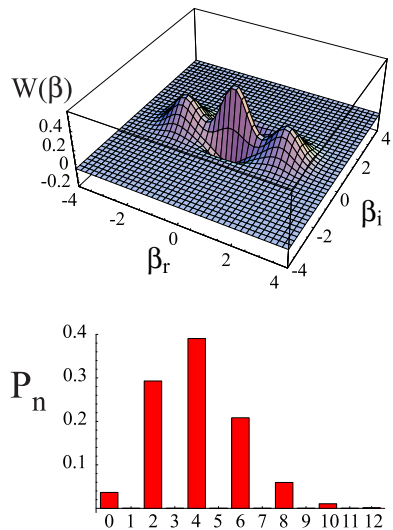}
\caption{\label{Fig1} (Color on line) Wigner function $W(\beta)$
and number probability distribution $P_n$ for the state given by
Eq.~(\ref{eq:inistate}) with $\alpha = 2$.}
\end{figure}

\subsection{Zeno and anti-Zeno dynamics of Schr\"{o}dinger cat states: the Wigner function}
\begin{figure}
\centering
\includegraphics[scale=0.3]{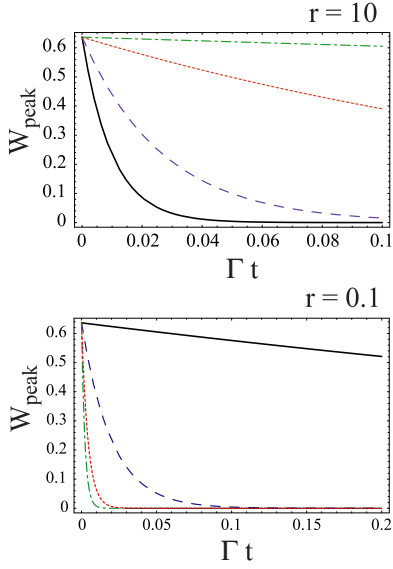}
\caption{\label{Fig2}(Color online) Dynamics of the Wigner
function peak for a high $T$ Ohmic reservoir, and for an initial
even coherent state with $\alpha=2$, in correspondence of the
parameters $r=10$ (figure above) and $r=0.1$ (figure below). In
both cases the solid (black) line represents the Markovian
dynamics in absence of shuttering. The dashed (blue) line
corresponds to $\omega_c \tau = 1$, the dotted (red) line
corresponds to $\omega_c \tau = 0.1$ and the dash-dotted (green)
line corresponds to $\omega_c \tau = 0.01$.}
\end{figure}

\begin{figure*}
\centering
\includegraphics[scale=0.7]{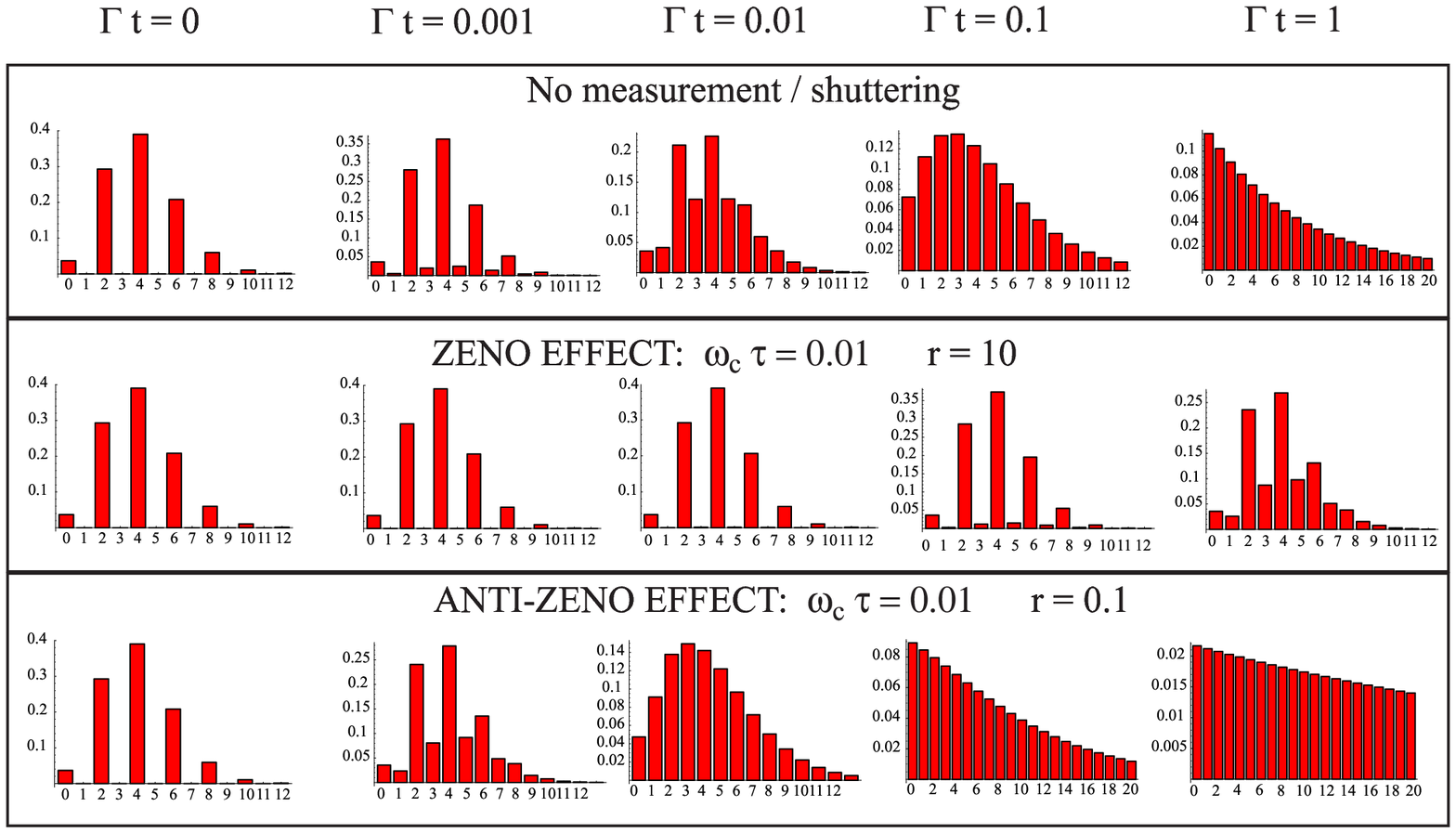}
\caption{\label{Fig3} (Color online) Number probability
distributions at different times $t$ for the case of no
measurements/shuttering (upper row), for the case of
measurments/shuttering with period $\omega_c \tau =0.01$ and
$r=10$ (middle row) or $r=0.1$ (lower row ), respectively.}
\end{figure*}

We now focus on the shuttered reservoir scenario, since we believe
that the realization of an experiment in this context might be
already within the grasp of the experimentalists. The time
evolution of the Wigner function can be calculated in two
equivalent ways. One can either use the QCF recursive solution
given by Eq.~(\ref{eq:recsol}), since the Wigner function is the
Fourier transform of the QCF,
\begin{eqnarray}
W(\beta)= \frac{1}{\pi^2} \int_{-\infty}^{\infty} d^2 \xi
\chi(\xi) \exp \left( \beta \xi^* - \beta^* \xi  \right).
\end{eqnarray}
Or, alternatively, one can use the Markovian solution for the
Wigner function corresponding to the master
equation~(\ref{Eq:RecME2}) [See, e.g., Ref.~\cite{Buzek92}], and
replace the Markovian decay coefficients with $\gamma_{\pm
1}(\tau)$. The analytical expression for the Wigner function at
time $t$, for a system initially prepared in the state given by
Eq.~(\ref{eq:inistate}) and subjected to the shuttered reservoir,
is the following
\begin{eqnarray}
W(\beta ,t) =  W^{(\alpha)}(\beta ,t) +  W^{(-\alpha)}(\beta ,t) +
W_I(\beta ,t),
\end{eqnarray}
with
\begin{eqnarray}
W^{\pm \alpha}(\beta , t)&=& \frac{2 {\cal N}}{\pi \left( 2 a_t+1
\right)} \exp \left( - \frac{2 \beta_i^2}{2
a_t+1}\right) \nonumber \\
&\times& \exp \left[ - \frac{2\left( \beta_r \mp e^{-b_{\tau} t/2}
\alpha \right)^2}{2 a_t+1} \right], \label{Wmix} \\
W_I(\beta ,t) &=& \frac{2 {\cal N}}{\pi \left( 2 a_t+1 \right)}
\exp \left( - \frac{2 |\beta|^2}{2
a_t+1}\right) \nonumber \\
&\times& \exp \left[ -2 \left( 1-\frac{e^{-b_{\tau} t} }{2
a_t+1}\right) \alpha^2 \right] \nonumber \\
&\times& \cos \left( \frac{4 e^{- b_{\tau} t/2}}{2a_t+1} \alpha
\beta_i \right).
\end{eqnarray}
In the previous equations $\beta_r$ and $\beta_i$ are the real and
imaginary part of $\beta$, respectively, and the functions $a_t$
and $b_{\tau}$  are defined as
\begin{eqnarray}
a_t &=& \frac{\gamma_{-1}(\tau)}{b_{\tau}} \left[ 1- e^{-b_{\tau}
t } \right], \label{eq:at}\\
b_{\tau} &=& \gamma_1(\tau)-\gamma_{-1}(\tau). \label{eq:btau}
\end{eqnarray}
We now look at the decay of the interference peak of the Wigner
function at $\beta = (0,0)$, defined as
\begin{eqnarray}
W_{\rm peak}(t)=W_I(\beta,t)|_{\beta=(0,0)}.
\end{eqnarray}
For a high $T$ reservoir, $W_{\rm peak}(t)$, during the initial
moments of the evolution, can be approximated as follows
\begin{eqnarray}
W_{\rm peak}(t) &\simeq& \frac{4 {\cal N}}{\pi} \exp \left[ -2
\gamma_{-1}(\tau)\left( 1+2 \alpha^2 \right)t\right].
\label{wpeak}
\end{eqnarray}
We consider a reservoir with an Ohmic spectral density with
Lorentzian cutoff \cite{Weiss99a}
\begin{eqnarray}
J(\omega)= \frac{2  \omega}{\pi} \
\frac{\omega_c^2}{\omega_c^2+\omega^2}, \label{Eq:SpecDen}
\end{eqnarray}
with $\omega_c$ the cutoff frequency. We note, incidentally, that
the reservoir correlation time $\tau_R$ is given, in this case, by
the inverse of the cutoff frequency. Under these conditions we
obtain the following analytic expression
\begin{eqnarray}
\gamma_{-1}(\tau) &=& \frac{\Gamma N(\omega_0)}{\tau \omega_c}
\Big\{ \omega_c \tau + \frac{1-r^2}{1+r^2}\left[ 1-e^{-\omega_c
\tau} \cos (\omega_0 \tau) \right] \nonumber \\
&-& \frac{2r}{1+r^2}e^{-\omega_c \tau} \sin (\omega_0 \tau)\Big\},
\label{gmeno1}
\end{eqnarray}
where
\begin{eqnarray}
\Gamma = 2 g^2 \frac{r^2}{r^2+1} \omega_0,
\end{eqnarray}
$\omega_0$ is the frequency of the system oscillator, and
$r=\omega_c/\omega_0$. As we have noted already in
Refs.~\cite{Maniscalco06a,Piilo07a}, the occurrence of Zeno or
anti-Zeno effects strongly depends on the value of the parameter
$r$. Values of $r$ smaller than unity indicate that the frequency
of the system oscillator is \lq\lq detuned\rq\rq~from the spectrum
of the reservoir, while values of $r$ greater than unity indicate
an overlap between the reservoir frequency spectrum and
$\omega_0$. Inserting Eq.~(\ref{gmeno1}) into Eq.~(\ref{wpeak})
one obtains the analytic expression for the time evolution of the
peak of the Wigner function for the shuttered reservoir case.

We notice that, when the shuttering period $\tau$ is much longer
than the reservoir correlation time, i.e. $\omega_c \tau \gg 1$,
one recovers the Markovian expression for the Wigner function peak
dynamics in absence of shuttering. Stated another way, as we have
already mentioned in this paper, for $\omega_c \tau \gg 1$ the
shuttering does not affect the dynamics. On the other hand, for
values of $\tau$ such that $\omega_c \tau \le 1$, one observes a
change in the Wigner peak dynamics depending on the behavior of
the coefficient $\gamma_{-1}(\tau)$, given by Eq.~(\ref{gmeno1}).

In Fig.~2 we compare the Markovian dynamics in absence of
shuttering and in presence of shuttering for $\omega_c \tau =
0.1$. For $r = 10$, i.e., in the case of a \lq\lq
resonant\rq\rq~reservoir the decay of the peak of the Wigner
function, indicating the passage from a quantum superposition to a
statistical mixture, is inhibited by the shuttering events, and
therefore the Schr\"{o}dinger cat lives longer. This is a
manifestation of the quantum Zeno effect. On the other hand, for
$r=0.1$, i.e., in the case of \lq\lq out of
resonance\rq\rq~reservoir, the peak of the Wigner function decays
faster, indicating a rapid passage from a quantum superposition to
a statistical mixture of the coherent states $\vert \alpha
\rangle$ and $\vert - \alpha \rangle$. This is a manifestation of
the anti-Zeno effect. Summarizing, by manipulating the parameters
of the engineered reservoir one can control the border between the
quantum and the classical worlds.

\subsection{Zeno and anti-Zeno dynamics of Schr\"{o}dinger cat states: the number probability
distribution}\label{sec:Pn}

We now consider the time evolution of the number probability
distribution. The dynamics of this quantity in presence of an
artificial shuttered reservoir coincides, as we have noticed, with
the dynamics of a system interacting with a natural environment in
presence of non-selective energy measurements. Therefore the
results here illustrated apply to both scenarios. Moreover, as it
is shown in Refs.~\cite{Intravaia03a,Grabert88a}, the expressions
for those observables which are diagonal in the Fock state basis
does not depend on the secular approximation performed to derive
the master equation (\ref{Eq:MQbm}), therefore the results
presented in this section go beyond the secular approximation too.

Following the lines of the Markovian derivation we solve directly
Eq.~(\ref{eq:Pn}) and obtain
\begin{eqnarray}
P_n(t) &=& \frac{2 {\cal N} e^{- \alpha^2}}{a_t+1}
\sum_{j=0}^l \frac{l!}{j![(l-j)!]^2} \left[\frac{a_t}{a_t+1}\right]^j \nonumber \\
&\times& \left[ \frac{\alpha^2 e^{- b_{\tau} t } }{\left( a_t+1
\right)^2}\right]^{l-j} \!\!\!\!\!\Big\{ \exp \left[ \left( 1-
\frac{e^{-b_{\tau} t}}{a_t+1}\right)\alpha^2\right] \nonumber \\
&+& (-1)^{l-j} \exp \left[ -\left( 1- \frac{e^{-b_{\tau}
t}}{a_t+1}\right)\alpha^2\right] \Big\},
\end{eqnarray}
with $a_t$ and $b_{\tau}$ given by
Eqs.~(\ref{eq:at})-(\ref{eq:btau}), respectively. The quantity
$a_t$ is the difference between the mean quantum number of the
system oscillator at time $t$, whose dynamics is studied in
Ref.~\cite{Piilo07a}, and the initial mean number of excitations.

We consider once more the case of a high $T$ reservoir with Ohmic
spectral density, as given by Eq.~(\ref{Eq:SpecDen}). In Fig.~3 we
plot the behavior of the number probability distribution. These
plots confirm the results of the previous section. Also by looking
at the number probability distribution, indeed, one sees that the
quantumness of the initial state, indicated by the absence of the
odd number components, can be prolonged or shortened in time,
compared to the situation in which no shuttering/measurements are
present.

\section{Conclusions}
In this paper we have investigated the dynamics of a damped
harmonic oscillator subjected to a series of non-selective energy
measurements performed at time intervals $\tau$. We have assumed
that the system is prepared initially in a Schr\"{o}dinger cat
state, i.e., a quantum superposition of two distinguishable
coherent states. This state is strongly sensitive to decoherence
induced by the external environment. The larger is the \lq\lq
separation\rq\rq~between the two components of the superposition,
the faster is the decay into a statistical mixture of the two
components.

We have shown, however, that when $\tau$ is smaller or of the same
order of the reservoir correlation time $\tau_R$, then the passage
from a quantum superposition to a classical statistical mixture
may be controlled. In more detail one can inhibit or enhance the
\lq\lq life\rq\rq~of the Schr\"{o}dinger cat depending on some
system/reservoir parameters. This result, which is based on the
occurrence of the quantum Zeno or anti-Zeno effects, respectively,
opens new possibility for protecting very fragile states like the
Schr\"{o}dinger cat states from the destructive effects of the
external environment.

We have also proved that one can mimic the effect of nonselective
measurements by modulating the system-reservoir coupling constant.
In this case one can use appropriately engineered reservoirs to
implement experiments aimed at moving in a controlled way the
quantum-classical border via the Zeno and anti-Zeno effect.

\section*{Acknowledgements}
The authors acknowledge financial support from the Academy of
Finland (projects 108699, 115982, 115682), the Magnus Ehrnrooth
Foundation and the V\"{a}is\"{a}l\"{a} Foundation. Stimulating
discussions with Saverio Pascazio and Paolo Facchi are gratefully
acknowledged.

\end{document}